Interlayer exchange coupling in $Co_2FeAl_{0.5}Si_{0.5}/Cr/Co_2FeAl_{0.5}Si_{0.5}$ trilayers

T. Furubayashi,[1, a)] K. Kodama,[2] H. S. Goripati,[2] Y. K. Takahashi,[1] K. Inomata,[1] and K. Hono[1, 2]

[1] *National Institute for Materials Science, Tsukuba 305-0047, Japan*

[2] *Graduate School of Pure and Applied Sciences, University of Tsukuba, Tsukuba 305-0047, Japan*

Electronic mail: furubayashi.takao@nims.go.jp

Interlayer exchange couplings were examined for Co$_2$FeAl$_{0.5}$Si$_{0.5}$(CFAS)/Cr/CFAS trilayered films grown on MgO (001) single crystal and thermally oxidized Si substrates. The films were (001) epitaxial on MgO and (110) textured polycrystalline on SiO$_2$. Strong exchange couplings were observed for the films with the 1.5 nm thick Cr spacer layer. A 90° coupling is dominant in the (001) epitaxial film. In contrast, an antiparallel coupling exists in the polycrystalline one. The relationship of interlayer couplings with the structure is discussed.

Much attention has been given to the problem of interlayer exchange couplings in multilayered films, especially multilayers consisting of half metallic full-Heusler alloys that have exhibited interesting results. Ambrose *et al.* [1] reported that, in Co$_2$MnGe(CMG)/V/CMG and CMG/Mn/CMG trilayers, the magnetizations of two CMG layers couple at 90° to each other, while the CMG/Cr/CMG films exhibited no significant coupling. Recently, Wang *el al.* [2] showed that (001) epitaxial Co$_2$MnSi(CMS)/Cr/CMS films show very strong 90° interlayer exchange couplings. Namely, when the energy of the exchange coupling is expressed by,

$$E_J = -J_1 \cos\theta - J_2 \cos^2\theta \qquad (1)$$

where $\theta$ is the relative angle between the magnetization directions of the two layers, the biquadratic term $J_2$ is dominant while the bilinear term $J_1$, which gives antiparallel couplings, is negligibly small. As well known, antiparallel couplings have been widely observed in similar systems with a Cr spacer layer, like Fe/Cr/Fe [3-5]. The mechanism of such strong 90° coupling is still unclear. It is of much interest to see if such a property is observed commonly in Heusler alloys or half metals, since half metallic properties of CMS have been shown theoretically [6-8] and by tunneling magnetoresistance (TMR) of magnetic tunneling junctions (MTJ) consisting of CMS [9]. To obtain further insight, we have examined trilayers consisting of a Heusler alloy Co$_2$FeAl$_{0.5}$Si$_{0.5}$ (CFAS). The half-metallic nature of CFAS was predicted theoretically.[10, 11] Also experimentally, it has been shown that CFAS has high spin polarization.[12, 13] In this work, magnetic properties were examined for the trilayer structure of CFAS/Cr/CFAS.

Multilayered structures with CFAS Heusler alloy were grown by dc magnetron sputtering in an ultrahigh vacuum system with the base pressure lower than 5×10$^{-7}$ Pa. The films were deposited upon MgO (001) single crystalline substrates and thermally oxidized Si substrates at the same time. The substrates were kept at room temperature (RT) during the deposition. The CFAS films were deposited from the stoichiometric $Co_{0.5}Fe_{0.25}Al_{0.125}Si_{0.125}$ alloy target. The composition of the CFAS film was found to be $Co_{0.531}Fe_{0.256}Al_{0.108}Si_{0.105}$ by inductively coupled plasma analysis. The layer structure was Cr(30)/CFAS(20)/Cr(1.5 or 2.5)/CFAS(5 or 10)/Ta(5) from the bottom, where the numbers in the parentheses indicate the thickness in nm. After depositing the bottom CFAS(20) layer, the films were annealed at 673 K for improving the structural order in some samples.

The structure of the films was examined by x-ray diffraction (XRD) with the Cu-Kα radiation using a 4-axis diffractometer recorded at room temperature and transmission electron microscopy (TEM). Magnetization curves were recorded at room temperature (RT) using a vibrating sample magnetometer (VSM) with a magnetic field applied in the film plane. For epitaxial films prepared on MgO (001), a magnetic field was applied in the direction parallel to the [110] axis of CFAS, the easy direction.

Full-Heusler alloy has the composition of $X_2YZ$ in the $L2_1$ ordered structure. However, the partially disordered $B2$ phase with the mixing of Y and Z and the totally disordered $A2$ phase with the bcc structure also exist. Figure 1 shows XRD patterns obtained by conventional 2θ-θ scans for the films with the layered structure of Cr(30)/CFAS(20)/Cr(1.5)/CFAS(5)/Ta(5), where the bottom CFAS(20) layer was annealed at 673 K. As indicated in in Fig. 1(a), the (100) reflection of CFAS shows $B2$ ordering of CFAS deposited on MgO (001) single crystals. The (1/2 1/2 1/2) reflections, which indicate $L2_1$ ordering, were explored for this sample by using the 4-axis diffractometer but they were not observed. Layers of Cr and CFAS deposited on MgO (001) were found to show a strong (001) texture, suggesting an epitaxial growth in the

(001) direction. In contrast, the 2θ-θ scan for the sample deposited on SiO$_2$ in Fig. 1(b) showed the strong (110) reflection indicating the (110) texture.

The high resolution cross sectional TEM image of the sample on MgO (001), shown in Fig. 2, indicates the epitaxial growth of the Cr and the CFAS layers in the (001) direction. The crystal structure of the both CFAS layers was examined by nano-beam electron diffraction. The diffraction patterns in Fig 2 show that the bottom CFAS layer, annealed at 673 K was in the *B*2 structure, while the not-annealed top CFAS layer is in the disordered *A*2 structure. Cross sectional TEM images are shown in Fig. 3 for the sample deposited on SiO$_2$. Layered structures were confirmed by the observations of electron energy loss spectroscopy (EELS) mapping of Cr atoms. The nano-beam diffraction for the CFAS layers in the sample prepared on SiO$_2$ indicates the disordered *A*2 structure of CFAS as shown in Fig. 3(c). The diffraction also shows that the crystals are aligned with the (110) axis perpendicular to the plane, consistent with the observation by XRD.

Figure 4 shows the magnetization curves at room temperature for the samples with different preparation conditions, the results indicate non-ferromagnetic coupling between two CFAS layers through the 1.5 nm thick Cr layer. The samples shown in Fig 4(a) have the stacking of Cr(30)/CFAS(20)/Cr(1.5)/CFAS(5)Ta(5). For the sample prepared on MgO, we obtain $M_R/M_S = 0.81$, where $M_R$ is the remanent magnetization and $M_S$ is the saturation magnetization. This value is near 0.8 expected for the case that the two CFAS layers with the different layer thicknesses of 20 nm and 5 nm are magnetized in the directions with the angle of 90° to each other. This is qualitatively the same as observed in the CMS/Cr/CMS (001) trilayer films.[2] The sample with the same sequence deposited on SiO$_2$ shows different magnetization curves. It showed smaller $M_R/M_S$ of 0.58, which is near the value 0.6 expected for the case that the two CFAS layers are magnetized in the antiparallel directions. This result indicates that, in the polycrystalline CFAS/Cr/CFAS film prepared on SiO$_2$, a 180° coupling from the bilinear term exists.

In the samples in Fig 4(b), the bottom CFAS layer was annealed at 673 K with the same stacking as those in Fig. 4(a). The values of $M_R/M_S$ are mostly the same for the both substrates MgO and SiO$_2$. Thus, the interlayer couplings are qualitatively unchanged by annealing. We find, however, the increase of the saturation field $H_S$ by annealing. In the case of the sample on MgO, $H_S$ increases from 250 Oe to 430 Oe, indicating that the biquadratic coupling was enhanced by annealing the CFAS layer. The coupling constant $J_2$ in Eq. (1) was estimated to be -0.08 erg/cm$^2$ from $H_S$. The value is to be compared to -1.8 erg/cm$^2$ obtained in the CMS/Cr/CMS system.[2] The increase of $H_S$ from 500 Oe to 800 Oe was also observed in the samples on SiO$_2$ by annealing, suggesting the enhancement of the coupling. In the sample shown in Fig 4(c), the thickness of the top CFAS layer was increased to 10 nm. The values of $M_R/M_S$ were found to decrease correspondingly. Figure 4(d) shows the results of the films with a 2.5 nm thick Cr spacer layer. We find that the non-saturating behaviors in the samples with the 1.5 nm Cr spacer are not observed here. Thus, the exchange coupling decreased with increasing the Cr thickness to 2.5 nm.

We have shown that the interlayer couplings in the CFAS/Cr/CFAS trilayer films are quite different depending on the substrate, MgO or SiO$_2$. There are two possible explanations for this difference, both structural in origin. One is the degree of the chemical ordering of the CFAS Heusler alloy layer. The $B$2 ordering was found in the CFAS layers epitaxially grown on the MgO (001) substrate, while the polycrystalline

CFAS layers on SiO$_2$ were in the disordered *A*2 structure. High spin polarization of CFAS attained by improving the structural ordering may be the origin of the of the 90° coupling. Another possible reason is the difference of the crystalline orientation, the (001) or the (110) texture of the CFAS and the Cr layers. The interlayer couplings might depend on the electronic structure at the interface of Cr and CFAS for different crystalline orientations. At the present, however, definitive conclusion has not yet obtained.

In summary, interlayer exchange couplings were examined for CFAS/Cr/CFAS trilayer films. Strong couplings were observed for the films with the 1.5 nm thick Cr interlayer. A 90° coupling is dominant in the *B*2 ordered (001) epitaxial film prepared on MgO (001) substrates, similar to the CMS/Cr/CMS system. In contrast, an antiparallel coupling exists in the *A*2 polycrystalline films with the (110) texture. Thus, the interlayer couplings in this system strongly depend on the crystal orientation of the layers and possibly on the chemical ordering as well. The mechanism of the interlayer couplings is expected to be clarified theoretically by accounting for the electronic structure of both Cr and CFAS.

KK and HSG thank the National Institute for Materials Science for providing a NIMS junior research assistantship. This work was partly supported by a Grant-in-Aid for Scientific Research (B) 20360322, a Grant-in-Aid for Scientific Research in Priority Area "Creation and control of spin current" 19048029, and the World Premier International Research Center Initiative (WPI Initiative) on Materials Nanoarchitronics, MEXT, Japan.

Fig.1: (color online) XRD patterns for the samples with the stacking Cr(30)/CFAS(20)/Cr(1.5)/CFAS(5)/Ta(5) (numbers are thickness in nm) from the bottom deposited on (a) MgO(001) and (b) thermally oxidized Si. The samples were annealed at 673 K for 1 h after depositing the bottom CFAS layer.

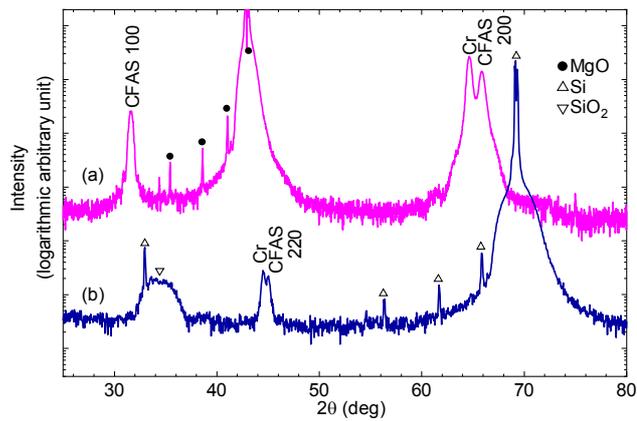

Fig. 2: (a) High resolution TEM image of the same sample as Fig. 1 deposited on MgO (001). Nano-beam diffraction patterns from (b) the top and (c) the bottom CFAS layers.

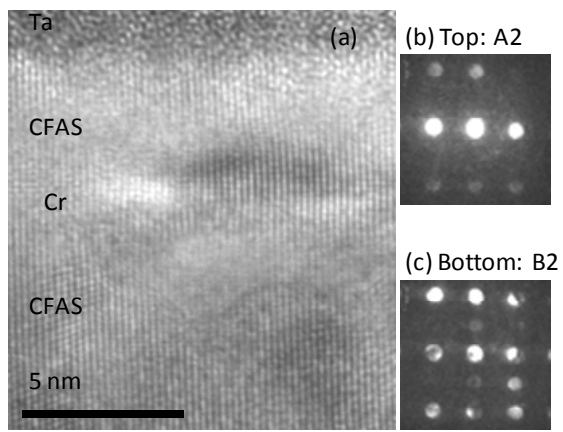

Fig.3: TEM images of the same sample with Fig. 1 deposited on thermally oxidized Si. (a) Bright field image, (b) Cr mapping image by EELS and (c) nano-beam diffraction pattern from the bottom CFAS layer.

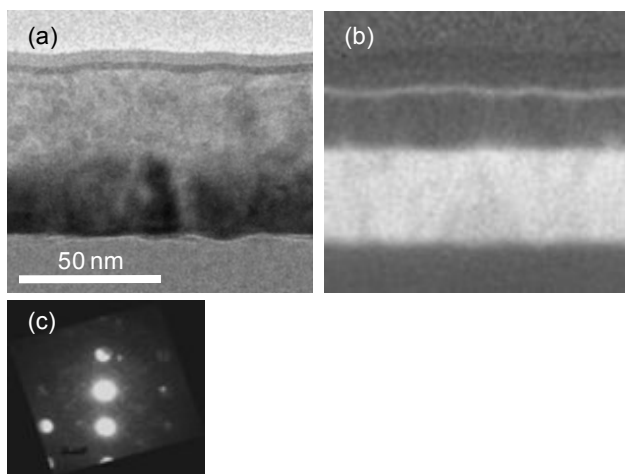

Fig. 4: (color online) Magnetization curves at RT for the samples with the stacking of (a) and (b) Cr(30)/CFAS(20)/Cr(1.5)/CFAS(5), (c) Cr(30)/CFAS(20)/Cr(1.5)/CFAS(10) and (d) Cr(30)/CFAS(20)/Cr(2.5)/CFAS(5) from the bottom. The samples shown in (b), (c) and (d) were annealed at 673 K after depositing the bottom CFAS layer. The close circles are for the samples deposited on MgO(001) and the triangles are for the samples deposited on thermally oxidized Si.

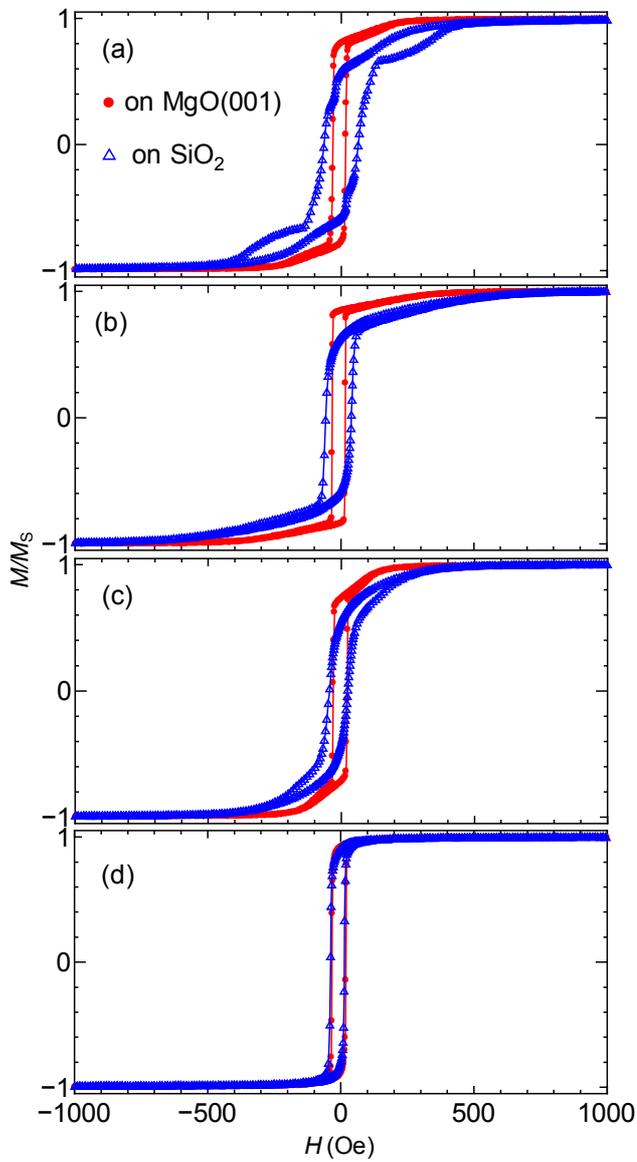